\documentstyle[psfig,volos]{article}
\begin{document}
%
%
\heading{%
 Type Ia Supernovae:\\
 Influence of the Progenitor on the Explosion}

\par\medskip\noindent
%
\author{%
Inma Dom\'\i nguez$^{1}$,
Peter H\"oflich$^{2,3}$,
Oscar Straniero$^{4}$
Craig Wheeler$^2$,
Friedrich-Karl Thielemann$^3$
}
\address{
Universidad de Granada, Granada, Spain
}
\address{
University of Texas, Austin, USA
}
\address{%
University of Basel, Basel, Switzerland
}
\address{%
Osservatorio Astronomico di Collurania, Teramo, Italy 
}
\begin{abstract}
The influence of the initial composition and structure of the exploding white dwarf on 
the nucleosynthesis and structure of Type Ia  Supernovae  has been studied. 
 The progenitor structures are based on detailed stellar evolutionary tracks for stars in the
mass range between 1 to 9 $M_\odot$ using the state of the art code FRANEC.
 The calculations of the thermonuclear explosions 
 are based on a set of delayed detonation models which give a good account of
the optical and infrared light curves  and of the spectral evolution.
  Our code solves the hydrodynamical equations explicitly by the piecewise parabolic method.
 Nuclear burning is taken into account using an extended network of 218
 nuclei.
 In principle, our calculations allow  the observed
 spectra and light curve to be linked to the progenitor. Moreover, our study
 is relevant to estimate potential evolution in the
 progenitor population at cosmological time scales.
\end{abstract}
\section{Explosion Models}
The influence of initial metallicity and C/O ratio on the light curves 
and spectra  has been studied for a set of delayed detonation models. 
The explosions are calculated using a one dimensional radiation-hydro 
 code, including nuclear network    
(\cite{HK96}  
 and references therein).   
For all the models, the WD mass is 1.4$M_\odot$, the central density of the WD is 
 2.6$\times 10^9$ g/cm$^3$ and the ratio of the  deflagration velocity 
 to the local sound speed is 0.03. 
 Table 1 shows basic parameters for the delayed detonation models:  
 these are, from  column 2 to 6, transition density (in g$/cm^3$), 
at which the deflagration is assumed to turn into a detonation, metallicity 
relative to solar, C/O ratio,   kinetic energy (in erg) and mass of $^{56}Ni$ 
 (in solar units). 
  The identification for the selected models is shown in 
 column 1, with DD21c  being the reference model.   

\begin{center}
\begin{tabular}{cccccc}
\multicolumn{6}{l}{{\bf Table 1.} 
Delayed Detonation Models  
} \\
\hline
\multicolumn{1}{c}
{Model} & 
\multicolumn{1}{c}
{$\rho_{tr}$} &
\multicolumn{1}{c}
{Z/Z$_\odot$} & 
\multicolumn{1}{c}
{ C/O} & 
\multicolumn{1}{c}
{ E$_{kin}$} & 
\multicolumn{1}{c}
{M$_{Ni}$ }  
\\
\hline
 {\bf DD21c} & 2.7$\times 10^7$ & 1.0 &  1.0 & 1.32$\times 10^{51}$ & 0.69 \\
 DD23c &2.7$\times 10^7$  & 1.0 & {\bf 0.66} & 1.18$\times 10^{51}$  & 0.59 \\
 DD24c & 2.7$\times 10^7$ & {\bf 1/3} & 1.0 & 1.32$\times 10^{51}$ & 0.70  \\
 DD13c & {\bf 3.0$\times 10^7$} & 1.0 & 1.0 & 1.36$\times 10^{51}$ & 0.79  \\
\hline
\end{tabular}
\end{center}

The corresponding bolometric light curves  
  have been calculated 
 (\cite{HWT98} and references therein).   
A decrease in  C implies a decrease in $^{56}Ni$ and
 consequently a decrease in the luminosity at later times  
(instantaneous deposition of the radioactive energy). However,  the maximum 
luminosity is greater because the smaller  kinetic energy causes a 
lower expansion rate and more of the stored energy contributes to the 
luminosity. Both effects cause a steeper decline rate after maximum light. 
Moreover if   
C is decreased, the energy release during nuclear burning is also lowered  and 
 the transition 
 density is reached later,  resulting in a larger pre-expansion 
 of the outer layers. In a first approximation, this is equivalent 
 to a reduction of the transition density. 
Lower expansion velocities and a  narrower region 
 dominated by Si are obtained in both cases.

The  
bolometric and monochromatic light curves are rather insensitive to 
the changes in metallicity. As  shown in Table 1, both the $^{56}Ni$ 
mass and the kinetic energy are close to the values obtained for the 
 {\it reference} model, DD21c.
 Synthetic NLTE spectra are calculated (\cite{H95}, \cite{HWT98}  
 and references therein). 
 The changes due to different metallicities can be directly observed in the UV spectral region.  
 Lower metallicities  and thus higher $Y_e$ produce less  
  $^{54}Fe$ \cite{NTY84}, which  
 provides an important source  for the opacity. 
However, this will only affect external regions 
 (with $v_{exp}$ above 12000 km/s); 
  Moreover  2 or 3 weeks after maximum when the spectrum is formed in the inner regions it  becomes  
 insensitive to the initial metallicity of the WD.  
\section{Evolutionary Models}
 Evolutionary tracks for a full set of intermediate mass 
 star models have been obtained using the 
code FRANEC \cite{Str98} and \cite{Dom98}.
 A full 
description of the updated  
input physics can be found in \cite{SCL97}. 
  The main problem related to the evolution 
of intermediate mass stars is the treatment  
 of turbulent convection. In this work the extension of 
the convective regions is fixed by the Schwarzschild criterion 
(without mechanical overshooting). Semiconvection during 
central helium burning is included, while the {\it breathing pulses}
 at the end 
of the central He burning phase are inhibited.  

  The main results are illustrated in Table 2,  where columns 1 to 9 show,  
respectively, the main sequence mass (in 
 solar units), the initial metallicity and helium, the mass of the 
C-O core at the thermal pulse phase (in solar units), the central mass 
 fraction of C, the C/O ratio within the core, the C/O ratio after 
 attaining  the Chandrasekhar mass (for the accreted matter C/O=1 
 is assumed), the source of the $^{12}C(\alpha,\gamma)^{16}O$ rate:  
  CF85 corresponds to Caughlan and Fowler \cite{CF85}, 
B-H96 and B-L96 
 to the upper and lower limits according to Buchmann \cite{Buc96} and 
 \cite{Buc97}  
 and   
  the numbers of thermal pulses calculated. Note that the C-O core chemical structure 
is taken at the last thermal pulse.   

The amount of C decreases as the WD progenitor mass increases 
or WD progenitor metallicity decreases  
 (see Table 2). 
The maximum difference obtained, once the C-O core is accreted to
the Chandrasekhar mass,  is about 18$\%$ (see Table 2 and Figure 1) for the 
maximum difference in progenitor masses 
and  
 about 5$\%$ for the maximum difference in progenitor metallicity 
(see Table 2 and Figure 2 {\it left}).

The assumed $^{12}C(\alpha,\gamma)^{16}O$ reaction rate constitutes a critical point  
(see Table 2 and Figure 2 {\it right}).  
 When central helium is depleted to 0.1 (mass 
fraction) the burning mainly occurs through this reaction, determining 
 the $^{12}C/^{16}O$ profile in the final WD. 
However, note  that the adoption of a particular 
 rate
should be done in accordance with the decision of the convection criterion. The treatment of convection in our code is consistent with a {\it high} 
reaction rate. 
\begin{figure}
\centerline
{\vbox{\psfig{figure=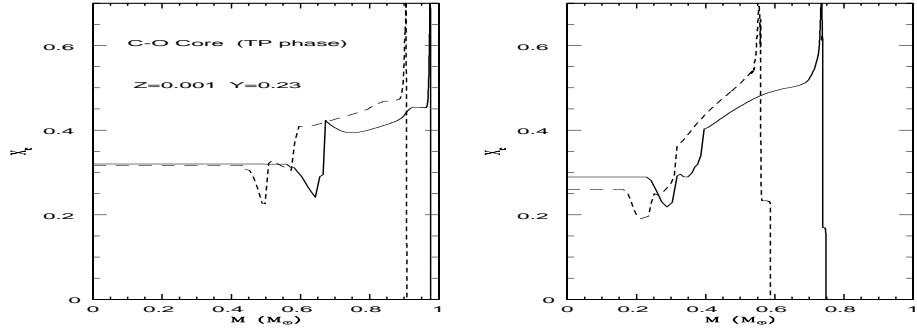,height=4.7cm,width=\textwidth}}}
\caption[]{\small
 Carbon abundance (mass fraction) in the C-O core  
for the  selected  masses (Z=0.001 and Y=0.23). {\it left}: 
6M$_\odot$ (solid) and 5M$_\odot$ (dashed).
 {\it right}: 3M$_\odot$ (solid) and 1.5M$_\odot$ (dashed).
}
\end{figure}

\begin{center}
\begin{tabular}{ccccccccc}
\multicolumn{9}{l}{{\bf Table 2.} 
Properties of the CO core at  the TP-AGB phase
} \\
\hline
\multicolumn{1}{c}
{M$_T$} &
\multicolumn{1}{c}
{Z} & 
\multicolumn{1}{c}
{Y} & 
\multicolumn{1}{c}
{M$_{CO}$} & 
\multicolumn{1}{c}
{ C$_{C}$} &
\multicolumn{1}{c}
{ C/O$_{core}$} & 
\multicolumn{1}{c}
{ C/O$_{Mch}$} & 
\multicolumn{1}{c}
{Rate} & 
\multicolumn{1}{c}
{TP (number)}  
\\
\hline
 1.5   & 0.001 & 0.23 & 0.560 & 0.260 & 0.533 & 0.781 & CF85 & 3 \\
 {\bf 3.0}   & 0.001 & 0.23 & {\bf 0.738} & {\bf 0.289} & {\bf 0.610} & {\bf 0.772} & CF85 & 2 \\
 5.0   & 0.001 & 0.23 & 0.907 & 0.316 & 0.563 & 0.691 & CF85 & 1 \\
 6.0   & 0.001 & 0.23 & 0.976 & 0.320 & 0.537 & 0.650 & CF85 & 1 \\
\hline
 3.0   & 0.0001 & 0.23 & 0.791 & 0.286 & 0.587 & 0.741 & CF85 & 1 \\
 3.0   & 0.001 & 0.28 & 0.815 & 0.297 & 0.623 & 0.760 & CF85 & 8 \\
 3.0   & 0.02 & 0.28 & 0.561 & 0.232 & 0.522 & 0.778 & CF85 & 3 \\
\hline
 3.0   & 0.001 & 0.23 & 0.750 & 0.279 & 0.570 & 0.742 & B-H96 & 6 \\
 3.0   & 0.001 & 0.23 & 0.727 & 0.654 & 2.314 & 1.525 & B-L96 & 5 \\
\hline
 3.0   & 0.001 & 0.23 & 0.759 & 0.289 & 0.621 & 0.773 & CF85 & 11 \\
\hline
\end{tabular}
\end{center}

\begin{figure}
\centerline
{\vbox{\psfig{figure=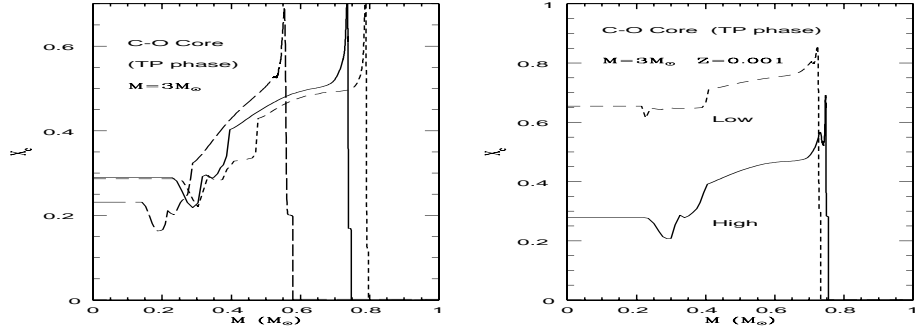,height=4.7cm,width=\textwidth}}}
\caption[]{\small
 Carbon abundance (mass fraction) in the C-O core.  
{\it left}:  Models with different  metallicities (3$M_\odot$): 
Z=0.0001, Y=0.23 (short dashed), Z=0.001, Y=0.23 (solid) and 
Z=0.02 Y=0.28 (long dashed). 
{\it right}: 
The high and low rates for the {\bf $^{12}C(\alpha,\gamma)^{16}O$}
 proposed  by L. Buchmann (1996, 1997) are adopted  
  (M=3M$_\odot$, Z=0.001 and Y=0.23).
}
\end{figure}

\section
{Supernovae at High Redshifts and Cosmology.}

The successful  detection of SNe Ia at high redshifts, currently around 
 96 SNe with redshifts ranging from 0.18 to 0.86 \cite{Per95},   
\cite{Per97}, \cite{Rie95} and \cite{Rie98},  
  has provided an exciting new tool for 
  cosmology research. An atonishing conclusion has been reached: results  
 are consistent with a  low matter density universe and a positive 
 cosmological constant. 
These results are based on two assumptions:  that SNe Ia are nearly 
 standard candles and that individual deviations can be detected 
using   
 a  local calibration of the brightness-decline relation.  
     
However, several observations indicate the presence of  
evolutionary effects: SNe Ia are  
 dimmer in elliptical galaxies and in the outer regions of spirals;    
in the central region of spirals,  both 
intrinsically brighter and dimmer SNe Ia occur, while    
 they  are relatively underrepresented in the bulges of spiral 
galaxies
 \cite{Bra96},   \cite{Ham96} and 
 \cite{WHW97}.

Time evolution is expected to produce several 
 effects (see \cite{HWT98} and \cite{DHWS98} for a complete 
 description). The most important effects directly related to  
 this study concern  the progenitor mass and metallicity.  
 At early epochs  
 the  mean progenitor mass is larger and 
 a smaller  C/O is expected   
(see Table 2).
 A smaller C/O increases 
  the peak-to-tail luminosity ratio. 
 Moreover, the  C/O ratio  
influences the  ignition conditions and the propagation of the 
burning front.  
 The initial  metallicity  determines  the electron-nucleon 
fraction of the outer layers and hence affects the products of nuclear 
 burning. The changes in the  spectrum with changes in  
metallicity  
have an important  effect on the  colours of SNe Ia at high redshifts, where they are shifted 
into other bands.

We obtained evolutionary effects  of the order of 0.15 to 
0.3 mag \cite{HWT98} and \cite{DHWS98}, 
 due to changes in the  
brightness-decline relation and in the spectra.  These  effects are small but are of the same order 
 as the brightness change (0.15 mag) imposed by cosmological deceleration.

\vfill
\end{document}